\documentclass[lettersize,journal]{IEEEtran}
\usepackage{amsmath,amsfonts,amssymb}
\usepackage{algorithmic}
\usepackage{algorithm}
\usepackage{array}
\usepackage[caption=false,font=normalsize,labelfont=sf,textfont=sf]{subfig}
\usepackage{textcomp}
\usepackage{stfloats}
\usepackage{url}
\usepackage{verbatim}
\usepackage{graphicx}
\usepackage{gensymb}
\usepackage{cite}
\usepackage{hhline}
\usepackage{abbreviations}
\usepackage[dvipsnames]{xcolor}
\usepackage{mathrsfs}  
\usepackage{caption}
\usepackage{subfig}
\usepackage{comment}

\usepackage[encapsulated]{CJK}
\usepackage{ucs}
\newcommand{\cntext}[1]{\begin{CJK}{UTF8}{gbsn}#1\end{CJK}}

\usepackage{hyperref}

\usepackage{booktabs}

\begin{document}

\title{On-sky performance of new 90~GHz detectors for the Cosmology Large Angular Scale Surveyor (CLASS)}
\author{Carolina N\'u\~nez, John W. Appel,  Michael K. Brewer, Sarah Marie Bruno, Rahul Datta,
Charles~L. Bennett, Ricardo Bustos, David~T. Chuss, Sumit Dahal, Kevin~L. Denis, Joseph Eimer, Thomas Essinger-Hileman, Kyle Helson, Tobias Marriage, Carolina~Morales P\'erez, Ivan~L. Padilla, Matthew~A. Petroff, Karwan Rostem, Duncan~J. Watts, Edward~J. Wollack, and Zhilei Xu~(\cntext{徐智磊})
\thanks{Carolina N\'u\~nez, John W. Appel, Charles~L. Bennett, Michael K. Brewer, Sarah Marie Bruno, Rahul Datta, Joseph Eimer, Tobias Marriage, Carolina~Morales P\'erez, and Ivan~L. Padilla are with the Department of Physics and Astronomy, Johns Hopkins University, Baltimore, MD 21218 USA.}
\thanks{Ricardo Bustos is with the Departamento de Ingenier\'{i}a El\'{e}ctrica, Universidad Cat\'{o}lica de la Sant\'{i}sima Concepci\'{o}n, Concepci\'{o}n, Chile.}
\thanks{Sumit Dahal, Kevin~L. Denis, Thomas Essinger-Hileman, Karwan Rosten, and Edward~J. Wollack are with the NASA 
Goddard Space Flight Center, Greenbelt, MD 20771, USA.}
\thanks{Kyle Helson is with the Center for Space Sciences and Technology, University of Maryland, Baltimore County, Baltimore, MD 21250, USA, and also with the NASA 
Goddard Space Flight Center, Greenbelt, MD 20771, USA.}
\thanks{Matthew~A. Petroff is with the Center for Astrophysics, Harvard \& Smithsonian, Cambridge, MA 02138, USA.}
\thanks{Duncan~J. Watts is with the Institute of Theoretical Astrophysics, University of Oslo, Oslo, Norway.}
\thanks{Zhilei Xu~(\cntext{徐智磊}) is with the MIT Kavli Institute, Massachusetts Institute of Technology, Cambridge, MA 02139, USA.}
}

\markboth{4EOr1A-01}%
{}

\maketitle

\begin{abstract}
The Cosmology Large Angular Scale Surveyor (CLASS) is a polarization-sensitive telescope array located at an altitude of 5,200 m in the Chilean Atacama Desert and designed to measure the polarized Cosmic Microwave Background (CMB) over large angular scales.
The CLASS array is currently observing with three telescopes covering four frequency bands: one at 40~GHz (Q); one at 90~GHz (W1); and one dichroic system at 150/220~GHz (HF).
During the austral winter of 2022, we upgraded the first 90~GHz telescope (W1) by replacing four of the seven focal plane modules.
These new modules contain detector wafers with an updated design, aimed at improving the optical efficiency and detector stability.
We present a description of the design changes and measurements of on-sky optical efficiencies derived from observations of Jupiter.

\end{abstract}


\section{Introduction}
\IEEEPARstart{T}{he} Cosmology Large Angular Scale Surveyor (CLASS) is a four-telescope polarization-sensitive array located on Cerro Toco at 5,200~m in the Chilean Atacama Desert.
The CLASS telescopes are designed to measure ``E-mode'' (even parity) and ``B-mode'' (odd parity) polarization patterns in the Cosmic Microwave Background (CMB) over large angular scales ($>1\degree$), with the goal of improving our understanding of inflation, reionization, and dark matter~\cite{Tom-overview},\cite{Harrington-overview}.

The CLASS array design consists of four telescopes: one at 40~GHz (Q), two at 90~GHz (W1 \& W2), and one dichroic system at 150/220~GHz (HF).  The CLASS array is currently observing with three telescopes across all four targeted frequency bands. The Q-band instrument was deployed in 2016, W1 was deployed in 2018, and HF was deployed in 2019.  CLASS focal planes consist of arrays of highly sensitive feedhorn-coupled transition-edge sensor (TES) bolometers 
that are voltage-biased to their superconducting transition critical temperatures of $\sim150\,\mathrm{mK}$.
The Q-band, W-band, and HF focal planes contain 72, 518, and 1020 TES bolometers, respectively. 
The CLASS TES bolometers provide the background-limited sensitivity required to achieve the experiment's science goals.
Characterization and on-sky performance of these focal planes are presented in \cite{Appel-Q, Dahal-W1, Dahal-HF, Dahal-multifrequency}.

During the austral winter of 2022, we upgraded the first 90~GHz telescope (W1) by replacing four (out of seven) of the focal plane modules.
These new modules contain detector wafers with an upgraded design, aimed at improving the optical efficiency and detector stability performance issues described in \cite{Dahal-multifrequency}.
In-lab testing and characterization of these detectors, including electrothermal parameters, bandpass measurements, and dark noise performance, are described in \cite{NunezSPIE}.
As a supplement to this previous work, this paper presents preliminary on-sky performance of these new detectors. In \S~\ref{sec:design}, we describe the upgraded wafer design.  In \S~\ref{sec:characterization}, we describe optical efficiency measurements derived from dedicated observations of Jupiter.

\section{Design}
\label{sec:design}

The CLASS 90~GHz detector wafers, which integrate 37 detector pixels, were fabricated at NASA Goddard Space Flight Center.
Each of the detector pixels consists of a symmetric planar ortho-mode transducer (OMT), which reads out two orthogonal linear polarizations to two TES bolometers.
Signals from opposing antenna probes are routed through a vialess crossover and a terminated vialess crossover, which symmetrizes the response between both OMT signal paths, and then coherently combined onto a single microstrip transmission line using the difference output of a Magic Tee, which transmits a single mode~\cite{U-Yen-magicT}.
On-chip filtering and micromachined silicon packaging define the signal bandpass~\cite{Crowe-choke}.
Finally, the signal from each of the two orthogonal polarizations is passed to a MoAu bilayer TES bolometer.
For a full description of the original CLASS 90~GHz wafer, see~\cite{Chuss-development, Denis-fabrication, Rostem-design}.

As summarized in \cite{NunezSPIE}, we present the design changes of the new 90~GHz detector wafers.  The updated  wafer includes the following main design changes to the original CLASS 90~GHz detector design:
\begin{enumerate}
    \item a simplified absorber that terminates power from the sky (brought in via a Nb microstrip) onto the TES island, with a resistive PdAu meander that consists of a stepped impedance transition from Nb to PdAu; 
    \item the addition of a direct normal-metal connection between the TES and the heat capacity element formed by the Pd, to effectively lump the electronic heat capacity into a single element; 
    \item the revision of the choke filter circuit design to extend onto the membrane’s diffusive bolometer legs.
    \item the addition of a revised absorber at the Magic Tee and at the terminated vialess crossover, realized as a lossy stepped impedance transition between Nb to PdAu that decreased the total meander length, device footprint, and sensitivity to detailed implementation.
\end{enumerate}

\begin{figure}[t!]
\centering
\includegraphics[width=2.5in]{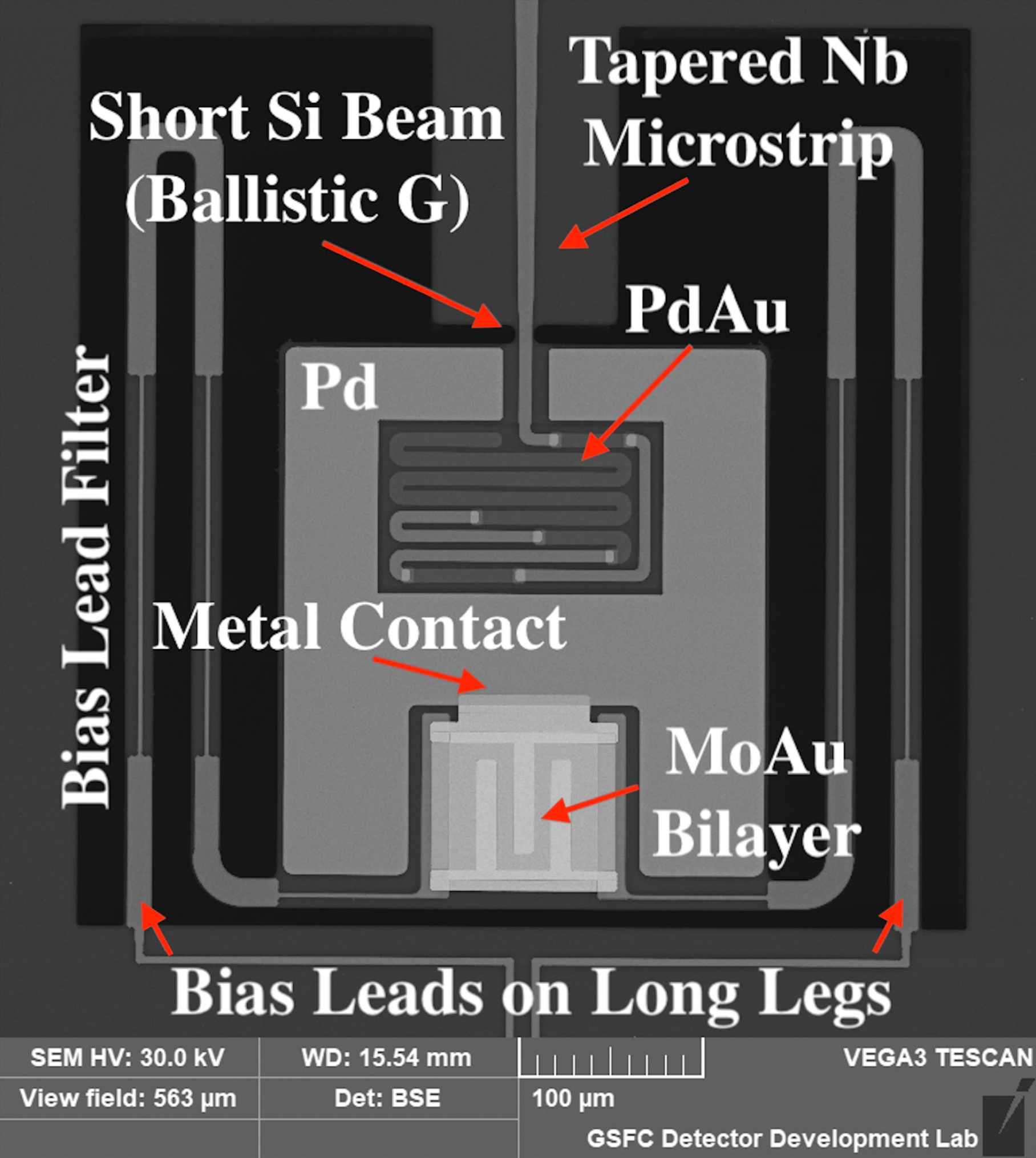}
\caption{Upgraded CLASS TES showing three primary design changes described in \S~\ref{sec:design}: 1) a simplified absorber with a resistive PdAu meander; 2) a direct normal-metal contact between the MoAu TES and the Pd heat capacity element; 3) the revised choke filter circuit implementation extending onto the membrane's diffusive bolometer legs. Power from the sky is brought in via the tapered Nb microstrip and terminates through the absorber onto the TES island.  The short Si beam sets the thermal conductance ($G$) and regulates the flow of power between the TES island and the bath. The MoAu bilayer determines the superconducting critical temperature ($T_\mathrm{c}$) of the TES. In-lab characterization of electrothermal parameters for the TES can be found in \cite{NunezSPIE}.}
\label{fig:TES}
\end{figure}

\begin{figure*}[!t]
    \centering
     \subfloat[]{\includegraphics[height=2.6in]{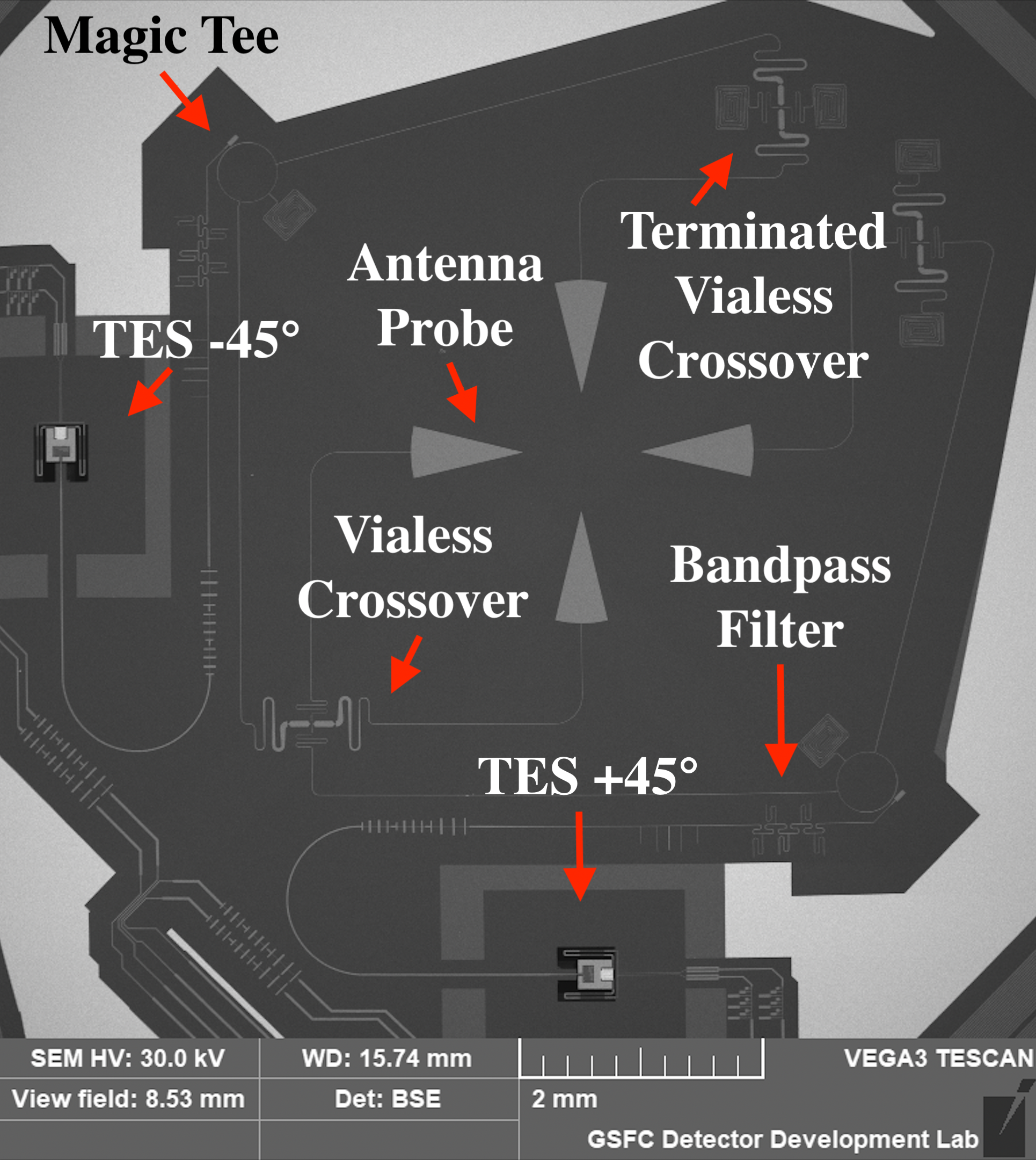}
     \label{subfig:pixel}}
     \hfil
     \subfloat[]{\includegraphics[height=2.6in]{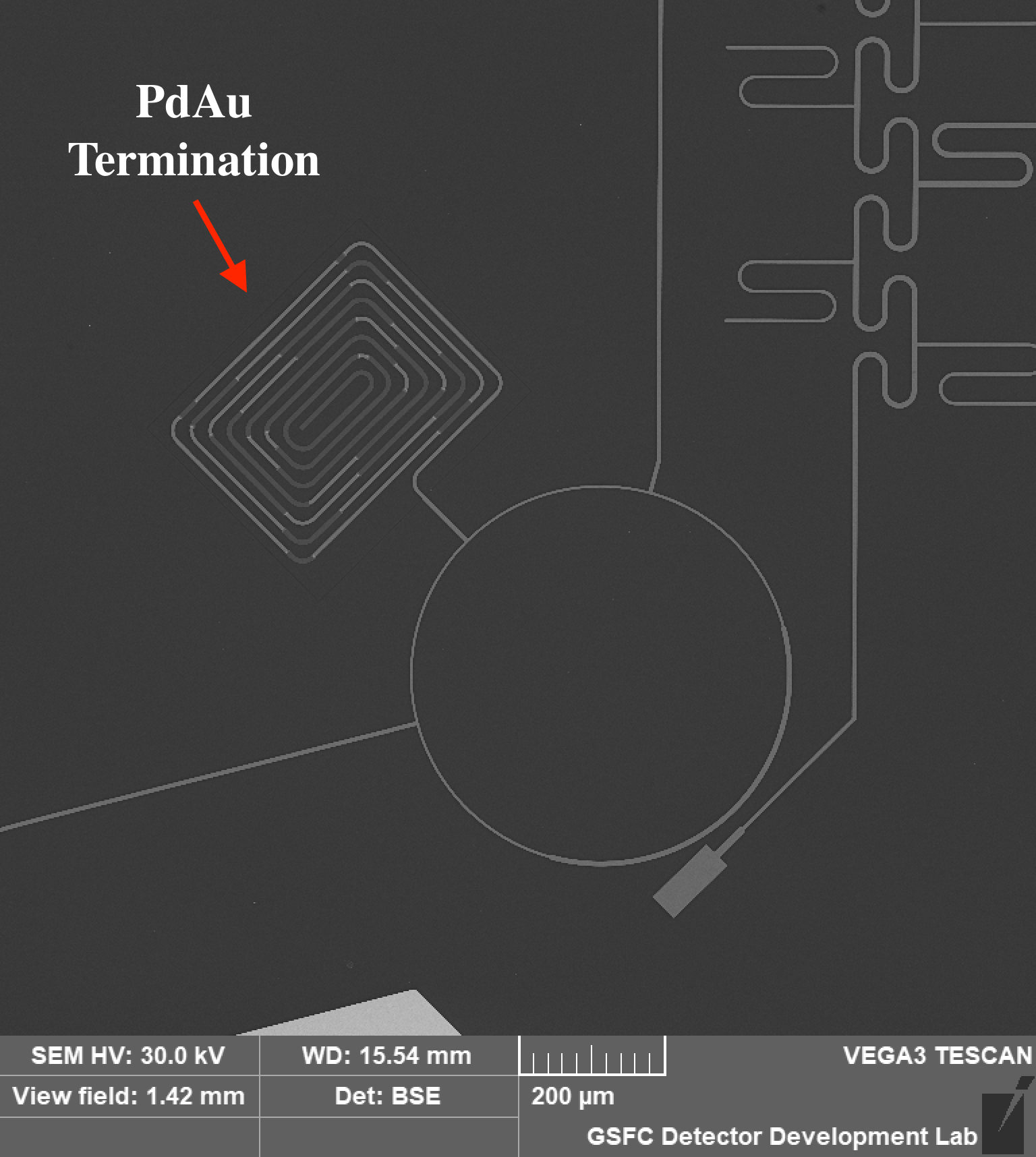}
     \label{subfig:magic_tee}}
     \hfil
     \subfloat[]{\includegraphics[height=2.6in]{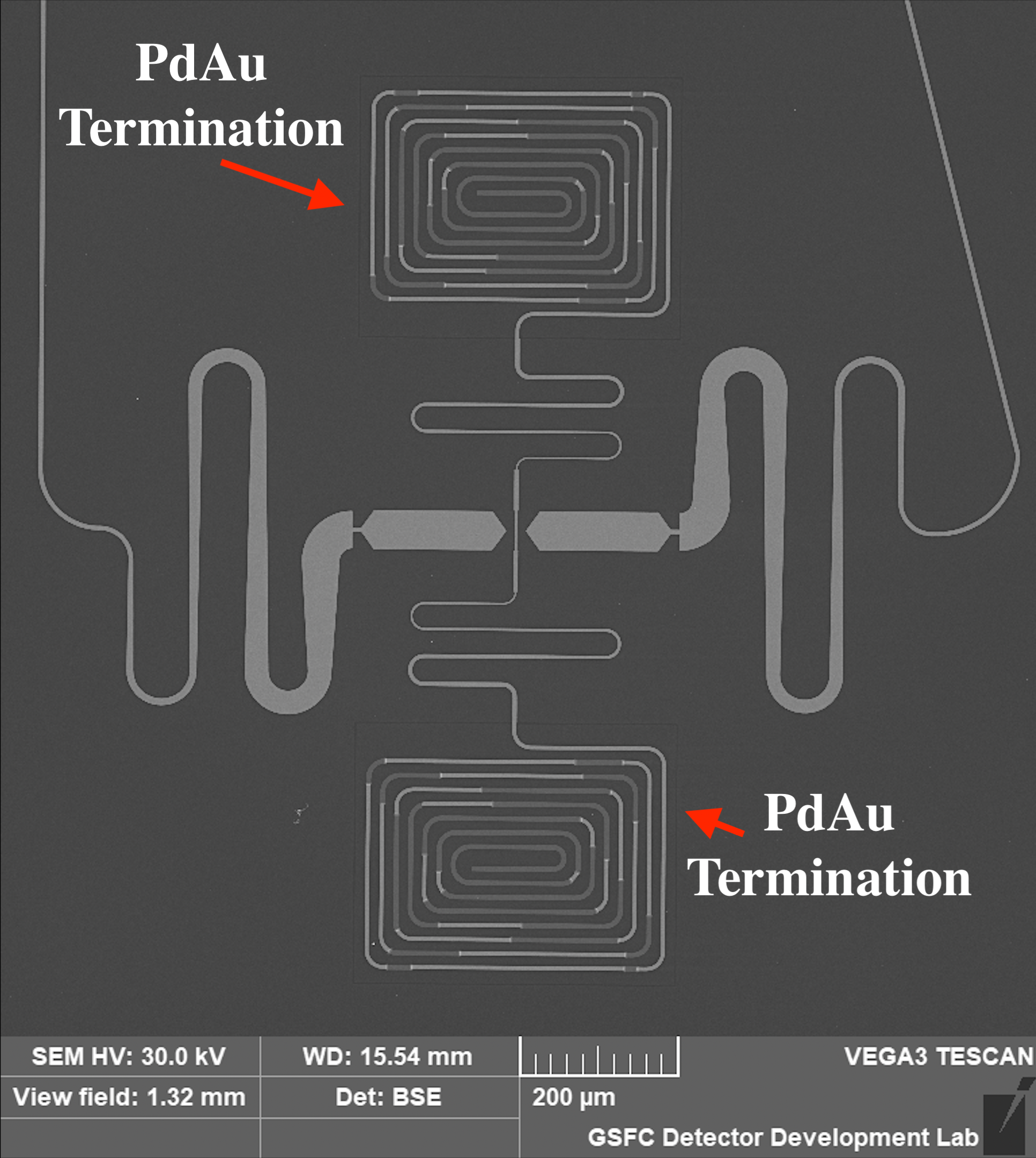}
     \label{subfig:crossover}}
     \caption{The upgraded CLASS 90~GHz detector pixel \textbf{(a)}, and zoom-ins showing the Magic Tee \textbf{(b)} and terminated vialess crossover \textbf{(c)} with the revised PdAu circuit termination. The new meandered PdAu termination was realized as a lossy stepped impedance transition between Nb to PdAu. This design strategy leads to a reduction of the total meander length, device footprint, and sensitivity to detailed implementation.}
     \label{fig:redesign}
\end{figure*}

These design changes were introduced in order to improve optical efficiency (1, 3) and stability of the TES transition (2), which were described in \cite{Dahal-multifrequency}.
The redesigned absorber at the Magic Tee and at the terminated crossovers (4) may have also improved the optical efficiency by achieving lower reflectance. The use of discrete interfaces along the length of absorber (rather than along the length of a taper) minimizes the uncertainty in the microwave propagation parameters arising from proximization \cite{Tinkham} at the superconducting and normal metal interfaces. 
The updated TES is shown in Fig.~\ref{fig:TES}. The full detector pixel and zoom-ins of the Magic Tee and terminated vialess crossover are shown in Fig.~\ref{fig:redesign}.

\section{On-Sky Performance: optical efficiencies}
\label{sec:characterization}

\begin{figure*}[!t]
    \centering
     \subfloat[]{\includegraphics[trim={0 0 0 0}, clip, height=2.25in]{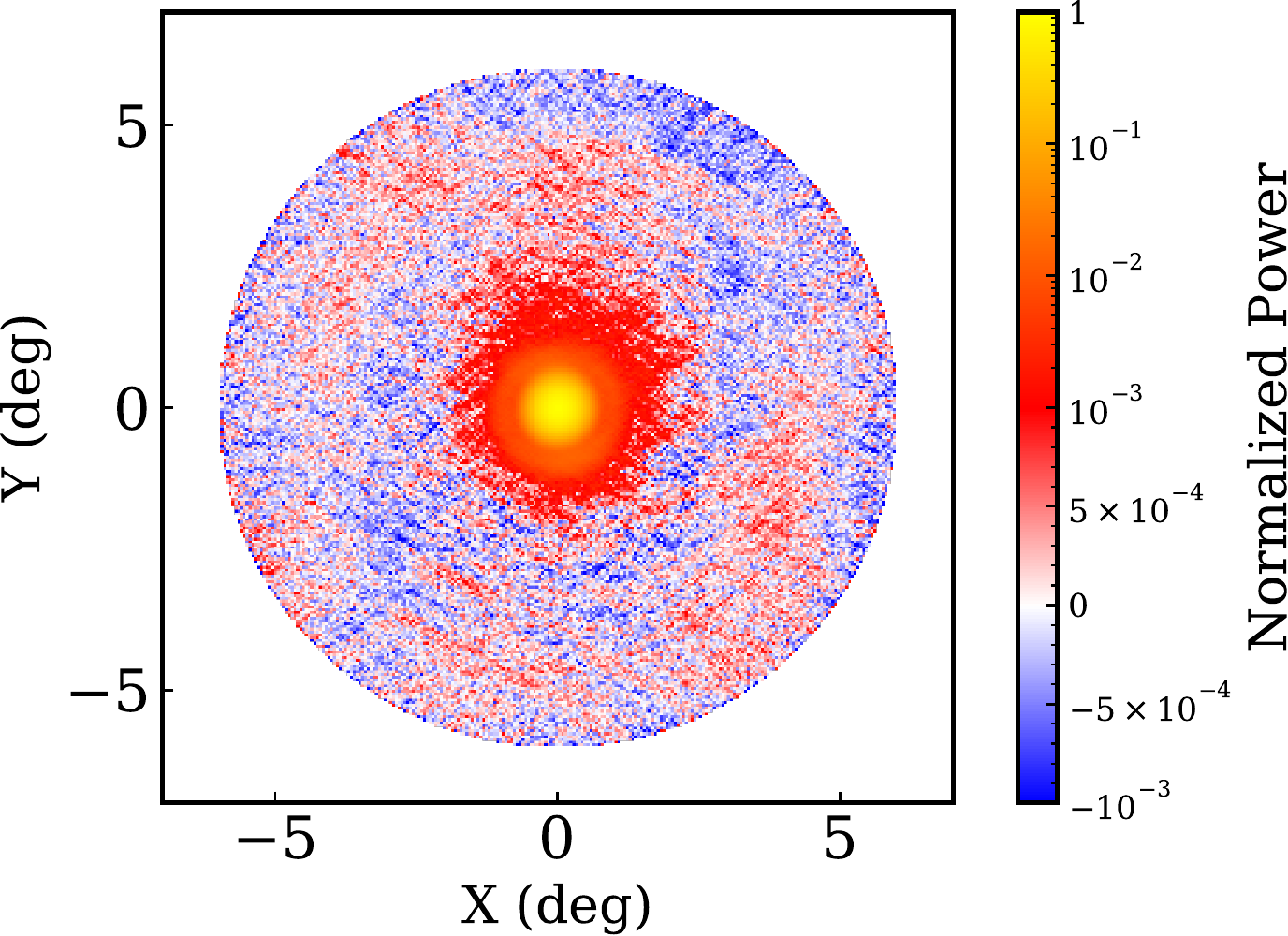}
     \label{subfig:jupiter_stack}}
     \hfil
     \subfloat[]{\includegraphics[height=2.25in]{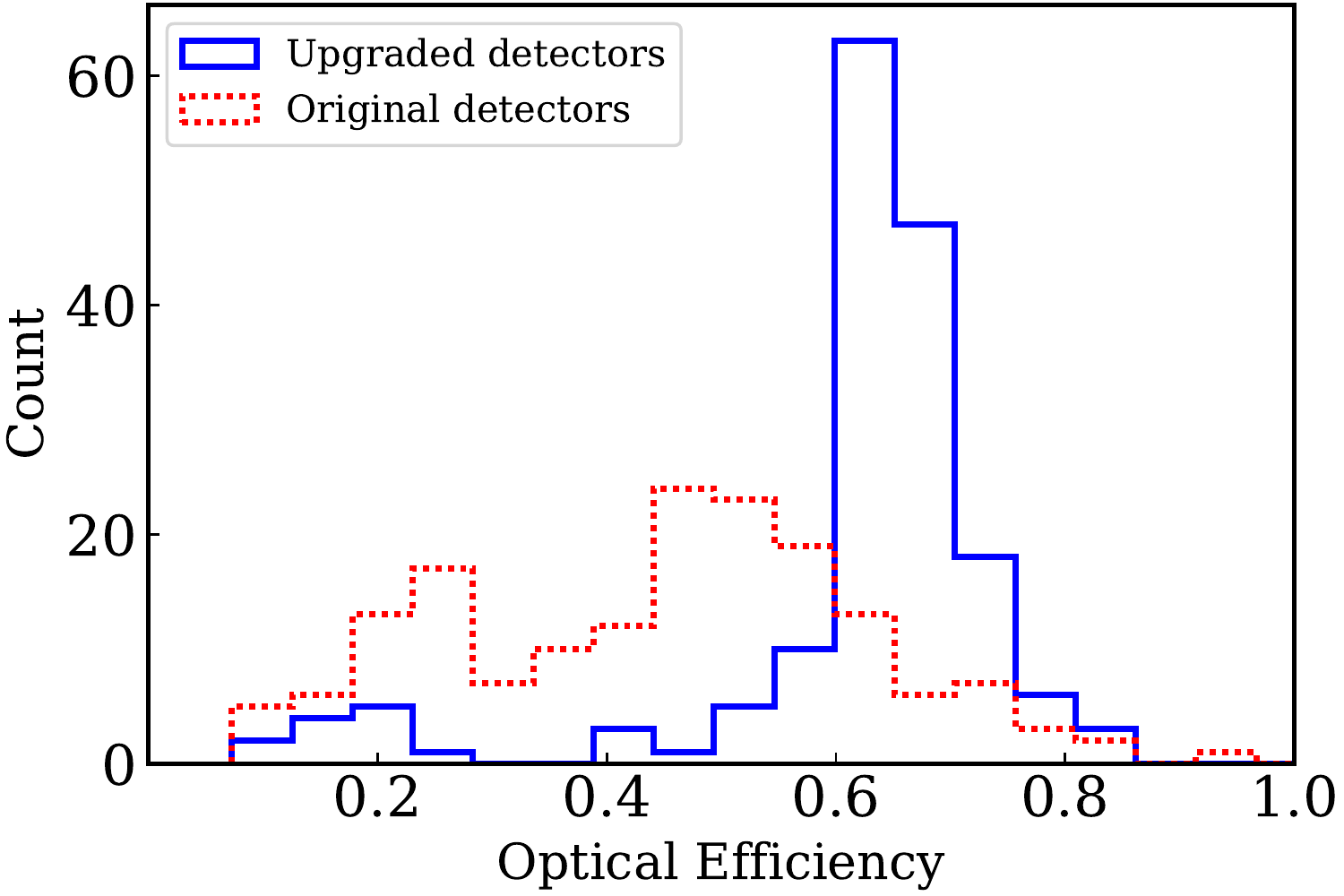}
     \label{subfig:efficiency_hist}}
     \caption{\textbf{(a)} A co-added map of Jupiter obtained from 85 observations of Jupiter with the upgraded 90~GHz focal plane.  The map is produced by co-adding signal from all 345 optical detectors, and is normalized to a peak amplitude of unity. The colorbar scale is linear from $-10^{-3}$ to $10^{-3}$ and logarithmic above $10^{-3}$. \textbf{(b)} A comparison of optical efficiencies between detectors on the currently fielded 90~GHz focal plane.  The focal plane contains four wafers with the upgraded design, and three wafers with the original design. 
     For the histogram of upgraded detectors, 13 detectors have optical efficiencies below 0.4.  Of these, 11 come from wafer 1 (following the labeling in Fig. 2d of \cite{NunezSPIE}.) We attribute the low optical efficiencies of wafer 1 to fabrication issues, which were improved for wafers 2--4. Overall, we see an improvement in optical efficiencies with respect to the previous focal plane.}
     \label{fig:optical_efficiency}
\end{figure*}

We measure the detector optical efficiencies via observations of Jupiter.  Specifically, we scan back and forth in azimuth at a constant elevation, while Jupiter rises or sets through the telescope beams. We can then measure the optical efficiency by comparing the observed and expected amplitude of Jupiter (i.e., the peak power received by the telescope when pointed directly at Jupiter).
In the Rayleigh-Jeans limit, the expected amplitude of Jupiter in this microwave frequency range is given by:

\begin{equation}
    A_\mathrm{J} = k_\mathrm{B} T_\mathrm{J} \Delta \nu B_{\mathrm{dil}}\,,
\end{equation}

\noindent where $k_\mathrm{B}$ is the Boltzmann constant, $T_\mathrm{J}$ is the Jupiter brightness temperature at 90~GHz (172.8~K) as reported by the WMAP team \cite{Jupiter}, $\Delta \nu$ is the observing bandwidth of 31~GHz, and $B_{\mathrm{dil}}$ is the beam dilution factor given by the ratio of the oblateness-corrected solid angle of Jupiter ($\Omega_\mathrm{J}$) and the convolution of the detector beam with Jupiter ($\Omega_\mathrm{B}$).
We can then obtain the absolute efficiency ($\eta$) by taking the ratio of the measured amplitude $A_{\mathrm{obs}}$ with the expected amplitude $A_\mathrm{J}$:

\begin{equation}
    \eta = \frac{A_{\mathrm{obs}}}{A_\mathrm{J}} = \frac{A_{\mathrm{obs}} \Omega_{\mathrm{B}}}{ k_\mathrm{B} \Delta\nu T_{\mathrm{J}} \Omega_{\mathrm{J}}}\,.
    \label{eqn:efficiency}
\end{equation}

\noindent To determine $A_{\mathrm{obs}}$, we do the following: initially, raw data from the Jupiter observations are converted from DAC units to units of power using the $I$-$V$ bin calibration method described in \cite{Appel-calibration} and low-pass filtered to remove the emission signal from the variable-delay polarization modulator (VPM) \cite{Katie-VPM, NathanVPM}.
The data are de-projected from the sky into the instrument frame, a $10\degree$ radius map centered on Jupiter is made for each detector, and a fit is done to a 2D elliptical Gaussian to derive the source amplitude.  This makes up the data that are used for each observation in the averaging.

During the averaging, each individual map is read in. After removing a baseline from each pass over the source in azimuth, the RMS noise is calculated from an annulus of the map well away from the source.
Since the solid angle of Jupiter varies over the course of the observations, we scale the data from the individual observations for each detector to a fiducial reference solid angle $\Omega_{\mathrm{ref}}$. For this we use an equatorial angular diameter of $48''$. The data are further scaled by $e^{\tau_i}$, where $\tau_i$ is the optical depth in the atmosphere, at each observation $i$, calculated from the zenith opacity as a function of precipitable water vapor (PWV) using the atmospheric model of \cite{PWV}, and then multiplied by the cosecant of the elevation of each detector to adjust the opacity at zenith to the opacity at the observing elevation.

Once all of the data for each detector are read in and processed, observations with an RMS noise of over three times the median for all observations are excluded. Out of a total of 85 observations, we retain an average of 75 observations per detector.  The surviving data are discretized onto a $6^\circ$ radius regular grid at $0.05^\circ$ resolution and a weighted average is calculated to yield the final averaged map for each detector:

\begin{equation}
\mathrm{Map}_{\mathrm{avg}} = \frac{\sum_{i=1}^{n_{\mathrm{obs}}} w_i (\Omega_{\mathrm{ref}} / \Omega_i) e^{\tau_i} \mathrm{Map}_i}{ \sum_{i=1}^{n_{\mathrm{obs}}} w_i} \,,
\end{equation}

\noindent where $\Omega_i$ is the solid angle obtained from the equatorial angular diameter of Jupiter at the time of each observation and $w_i$ is the signal-to-noise ratio calculated from the initial fitted amplitude and the RMS noise. The averaged map is then fit to a 2D elliptical Gaussian, yielding a peak amplitude $A_\mathrm{obs}$, and the beam solid angle $\Omega_B$ is measured by integrating the normalized map out to a radius of 
three times the full width at half maximum (FWHM) of the beam from the peak.  The efficiency can then be calculated, using $\Omega_{\mathrm{ref}}$ corrected for the oblateness of Jupiter, according to the method described in \cite{Weiland}, as $\Omega_\mathrm{J}$ in (\ref{eqn:efficiency}). Using the $2.974^\circ$ average observed latitude for the center of the disk of Jupiter yields $\Omega_\mathrm{J} = 3.978 \mathrm{x} 10^{-8}~\mathrm{sr}$.

Fig.~\ref{subfig:jupiter_stack} shows the observed map of Jupiter, composed of co-added maps from all 345 optical detectors.
In Fig.~\ref{subfig:efficiency_hist}, we show the distribution of optical efficiencies across the currently fielded focal plane.  The detectors with the updated design (four of the seven modules) are shown in blue, while the previous generation of detectors (three of the seven modules) are shown in red. We observe a marked improvement in the distribution of optical efficiencies for the detectors with the updated design. Furthermore, we note that the three wafers from the original design were the best-performing wafers of the original 90~GHz focal plane, i.e., the four upgraded detector wafers replaced the four least-performing wafers of the original focal plane.  Therefore, the distribution of the full original W1 focal plane favors even lower optical efficiencies than are shown for the original W1 detectors in Fig.~\ref{subfig:efficiency_hist}.
The optical efficiency distribution for the original 90~GHz focal plane is shown in \cite[Fig.~3]{Dahal-multifrequency}.

\section{Conclusion}
In conjunction with \cite{NunezSPIE}, we have provided in-lab characterization (electrothermal parameters, bandpasses, dark noise measurements) and on-sky performance (optical efficiencies) results for the new detectors of the CLASS 90~GHz focal plane.  These detectors obtained first light in the austral winter of 2022, and replaced four of the seven wafers of the original 90~GHz focal plane.  The detectors were redesigned with three primary changes to the TES aimed at improving optical efficiency and detector stability.  
In addition, changes to the terminations of the Magic Tee and terminated vialess crossover were implemented to reduce their reflectance, sensitivity to fabrication details, and the fidelity of the impedance match seen by the Magic Tee and crossover circuits.
We demonstrate improvements in optical efficiency between the former wafer design and current wafer design, by comparing expected vs. observed amplitude measurements of Jupiter.
A subsequent publication will provide further characterization and on-sky performance analysis of the upgraded 90~GHz focal plane.

\section*{Acknowledgment}
We acknowledge the National Science Foundation Division of Astronomical Sciences for their support of CLASS under Grant Numbers 0959349, 1429236, 1636634, 1654494, 2034400, and 2109311.
We thank Johns Hopkins University President R. Daniels and the Krieger School of Arts and Sciences Deans for their steadfast support of CLASS.
We further acknowledge the very generous support of Jim and Heather Murren (JHU A\&S '88), Matthew Polk (JHU A\&S Physics BS '71), David Nicholson, and Michael Bloomberg (JHU Engineering '64).
The CLASS project employs detector technology developed in collaboration between JHU and Goddard Space Flight Center under several previous and ongoing NASA grants. Detector development work at JHU was funded by NASA cooperative agreement 80NSSC19M0005.
Kyle Helson is supported by NASA under award number 80GSFC17M0002.
Zhilei Xu is supported by the Gordon and Betty Moore Foundation through grant GBMF5215 to the Massachusetts Institute of Technology.
We acknowledge scientific and engineering contributions from Max Abitbol, Fletcher Boone, David Carcamo, Lance Corbett, Ted Grunberg, Saianeesh Haridas, Jake Hassan, Connor Henley, Ben Keller, Lindsay Lowry, Nick Mehrle, Sasha Novak, Diva Parekh, Isu Ravi, Gary Rhodes, Daniel Swartz, Bingjie Wang, Qinan Wang, Tiffany Wei, Zi\'ang Yan, and Zhuo Zhang. We thank Miguel Angel D\'iaz, Jill Hanson, William Deysher, and Chantal Boisvert for logistical support.
We acknowledge productive collaboration with Dean Carpenter and the JHU Physical Sciences Machine Shop team.
CLASS is located in the Parque Astron\'omico Atacama in northern Chile under the auspices of the Agencia Nacional de Investigaci\'on y Desarrollo (ANID). 

\bibliography{main}
\bibliographystyle{IEEEtran}

\vfill

\end{document}